\documentclass[smallextended]{svjour3} 
\usepackage[square,numbers]{natbib}
\RequirePackage[colorlinks,citecolor=blue,urlcolor=blue]{hyperref}
\usepackage{amssymb,multirow}
\usepackage{amsfonts}
\usepackage{stmaryrd}
\usepackage{dsfont}

\usepackage{def_ju}

\usepackage{algorithm}
\usepackage{algorithmic}

\usepackage{ucs}
\usepackage[utf8x]{inputenc}
\usepackage{amsmath}
\usepackage{amsfonts}
\usepackage{amssymb}
\usepackage{fontenc}
\usepackage{graphicx}

\usepackage{bbm}
\usepackage{bm}
\RequirePackage[OT1]{fontenc}

\newcommand{\R}{\mathds{R}}
\newcommand{\E}{\mathds{E}}

\newcommand{\ddr}{\mathrm{d}}
\newcommand{\edr}{\mathrm{e}}

\def\simind{\stackrel{\mbox{\scriptsize{ind}}}{\sim}}
\def\simiid{\stackrel{\mbox{\scriptsize{iid}}}{\sim}}
\def\domainfn{(0,1)\times\mathbb{R}_+^*}
\newtheorem{thm}{\textsc{Theorem}}

\newtheorem{exe}{\textsc{Example}}
\allowdisplaybreaks

\newcommand*{\inlineequation}[2][]{%
  \begingroup
    \refstepcounter{equation}%
    \ifx\\#1\\%
    \else
      \label{#1}%
    \fi
    \relpenalty=10000 %
    \binoppenalty=10000 %
    \ensuremath{%
      #2%
    }%
    \hfill\@eqnnum
  \endgroup
}

\def\points{The points are connected by straight lines only for visual simplification.\xspace}
\def\samplefifty{The sample size on the $x$-axis in log scale runs from $n=50$ to $n=500$.\xspace}

\begin{document}

\title{Approximating predictive probabilities of Gibbs-type priors
}
%
%
%
%
%
%


\author{Julyan Arbel \and Stefano Favaro
}


\institute{
Julyan Arbel\at
Univ. Grenoble Alpes, Inria, CNRS, LJK, 38000 Grenoble, France\\
\email{julyan.arbel@inria.fr}
\and
Stefano Favaro \at
University of Torino and Collegio Carlo Alberto\\
Also  affiliated to IMATI-CNR  ``Enrico  Magenes" (Milan, Italy)\\
 \email{stefano.favaro@unito.it\\}  
}

\date{Received: date / Accepted: date}

\newcommand{\julyan}[1]{{\color{magenta}[Julyan: \rm{#1}]}}

\maketitle

\begin{abstract}
Gibbs-type random probability measures, or Gibbs-type priors, are arguably the most ``natural" generalization of the celebrated Dirichlet prior. Among them the two parameter Poisson--Dirichlet prior certainly stands out in terms of mathematical tractability and interpretability of its predictive probabilities, which made it the natural candidate in a plethora of applications. Given a random sample of size $n$ from an arbitrary Gibbs-type prior, we show that the corresponding predictive probabilities admit a large $n$ approximation, with an error term vanishing as $o(1/n)$, which maintains the same desirable features as the predictive probabilities of the two parameter Poisson--Dirichlet prior. Our result is illustrated through an extensive simulation study, which includes an application in the context of Bayesian nonparametric mixture modeling.
\keywords{Bayesian nonparametrics; first and second order asymptotic approximations; Gibbs-type prior; predictive probabilities; mixture modeling; normalized generalized Gamma prior; two parameter Poisson--Dirichlet prior}
\end{abstract}

%



\section{Introduction} 

Gibbs-type random probability measures, or Gibbs-type priors, are arguably the most ``natural" generalization of the Dirichlet process prior by Ferguson \cite{Fer(73)}. They 
have been first introduced in the seminal works of Pitman \cite{Pit(03)} and Gnedin and Pitman \cite{Gne(06)}, and their importance in Bayesian nonparametrics have been extensively discussed in Lijoi and Pr\"unster \cite{Lij(10)}, De Blasi et al. \cite{Deb(15)} and Bacallado et al. \cite{Bac(17)}. Gibbs-type priors have been widely used in the context of Bayesian nonparametric  inference for species sampling problems, where their mathematical tractably allowed to obtain explicit expressions for the posterior distributions of various population's features, and to predict features of additional unobservable samples. See, e.g., Lijoi et al. \cite{Lij(07)}, Lijoi et al. \cite{Lij(08)}, Favaro et al. \cite{Fav(09)}, Favaro et al. \cite{Fav(12)}, Bacallado 
et al. \cite{Bac(15)} and Arbel et al. \cite{Arb(15)}. The class of Gibbs-type priors has been also applied in the context of nonparametric mixture modeling, thus generalizing the celebrated Dirichlet process mixture model of Lo \cite{Lo(84)}. In particular, nonparametric mixture models based in Gibbs-type priors are characterized by a more flexible parameterization than Dirichlet process mixture model, thus allowing for a better control of the clustering behaviour. See, e.g., Ishwaran and James \cite{Ish(01)}, Lijoi et al. \cite{Lij(05)}, Lijoi et al. 
\cite{Lij(07a)}, Favaro and Walker \cite{Fav(13)} and Lomeli et al. \cite{Lom(15)}. Most recently, Gibbs-type priors  have been used in Bayesian nonparametric inference for ranked data (Caron et al. \cite{Car(14)}), sparse exchangeable random graphs and networks (Caron and Fox \cite{Car(15)} and Herlau \cite{Her(15)}), 
exchangeable feature allocations (e.g., Teh and G\"or\"ur \cite{Teh(10a)}, Broderick et al. \cite{Bro(13)}, Heaukulani and Roy \cite{Hea(15)}, Roy \cite{Roy(14)} and Battiston et al. 
\cite{Bat(16)}), reversible Markov chains (Bacallado et al. \cite{Bac(13)}), dynamic textual data (Chen et al. \cite{Che(12)} and Chen et al. \cite{Che(13)}), and bipartite 
graphs (Caron \cite{Car(12)}).

The definition of Gibbs-type random probability measures relies on the notion of $\alpha$-stable Poisson--Kingman model, first introduced by Pitman \cite{Pit(03)}. Specifically, let $(J_{i})_{i\geq1}$ be the  decreasing ordered jumps of an $\alpha$-stable subordinator, i.e. subordinator with L\'evy measure $\rho(\ddr x)=C_{\alpha}x^{-
\alpha-1}\ddr x$ for some constant $C_{\alpha}$, and let $P_{i}=J_{i}/T_{\alpha}$ with $T_{\alpha}=\sum_{i\geq1}J_{i}<+\infty$ almost surely; in particular $T_{\alpha}$ is 
a positive $\alpha$-stable random variable, and we denote its density function by $f_{\alpha}$. If $\text{PK}(\alpha;t)$ denotes the conditional distribution of $(P_{i})_{i\geq1}$ 
given $T_{\alpha}=t$, and if $T_{\alpha,h}$ is a random variable with density function $f_{T_{\alpha,h}}(t)=h(t)f_{\alpha}(t)$, for any nonnegative function $h$, then an $
\alpha$-stable Poisson--Kingman model is defined as the discrete random probability measure $P_{\alpha,h}=\sum_{i\geq1}P_{i,h}\delta_{X_{i}^{\ast}}$, where $
(P_{i,h})_{i\geq1}$ is distributed as $\int_{(0,+\infty)}\text{PK}(\alpha;t)f_{T_{\alpha,h}}(t)\ddr t$ and $(X_{i}^{\ast})_{i\geq1}$ are random variables, independent of $
(P_{i,h})_{i\geq1}$, and independent and identically distributed according to a nonatomic probability measure $\nu_{0}$. An $\alpha$-stable Poisson--Kingman model thus provides with a generalization of the normalized $\alpha$-stable process in Kingman \cite{Kin(75)}, which is recovered by setting $h=1$. According to the work of Gnedin and Pitman \cite{Gne(06)}, Gibbs-type random probability measures are defined as a class of (almost sure) discrete random probability measures indexed by a parameter $\alpha<1$ such that: i) for any $\alpha<0$ they are $M$-dimensional symmetric Dirichlet distribution, with $M$ being a nonnegative random variable on the set $\mathbb{N}$; ii) for $\alpha=0$ they coincide with the Dirichlet process; iii) for any $\alpha\in(0,1)$ they are $\alpha$-stable Poisson--Kingman models.

In this paper we focus on the predictive probabilities of Gibbs-type priors with $\alpha\in(0,1)$, i.e. the posterior expectation $\E[P_{\alpha,h}(\cdot)\,|\,\Xn]$, with $\Xn = (X_{1},\ldots,X_{n})$ being a random sample from $P_{\alpha,h}$. Due to the (almost sure) discreteness of the Gibbs-type random probability measure $P_{\alpha,h}$, we expect ties in a sample $\Xn$ from $P_{\alpha,h}$, that is $\Xn$ features $K_{n}=k_{n}\leq n$ distinct types, labelled by $X^{\ast}_{1},\ldots,X^{\ast}_{K_{n}}$, with corresponding frequencies $(N_{1},\ldots,N_{K_{n}})=(n_{1},\ldots,n_{k_{n}})$ such that $\sum_{1\leq i\leq k_{n}}n_{i}=n$. That is, the sample $\Xn$ induces a random partition of the set $\{1,\ldots,n\}$; see Pitman \cite{Pit(06)} for details on Gibbs-type random partitions. According to Pitman \cite{Pit(03)}, the predictive probabilities of $P_{\alpha,h}$ are
\begin{equation}\label{eq:pred_gibbs}
\text{Pr}[X_{n+1}\in\cdot\,|\,\Xn]=\frac{V_{n+1,k_{n}+1}}{V_{n,k_{n}}}\nu_{0}(\cdot)+\frac{V_{n+1,k_{n}}}{V_{n,k_{n}}}\sum_{i=1}^{k_{n}}(n_{i}-\alpha)\delta_{X^{\ast}_{i}}(\cdot)
\end{equation}
for $n\geq1$, where
\begin{equation}\label{eq:weight}
V_{n,k_{n}}=\frac{\alpha^{k_{n}}}{\Gamma(n-k_{n}\alpha)}\int_{0}^{+\infty}\int_{0}^{1}t^{-k_{n}\alpha}p^{n-k_{n}\alpha-1}h(t)f_{\alpha}((1-p)t)\ddr t\ddr p,
\end{equation}
with $\Gamma(\cdot)$ being the Gamma function. See, e.g., Pitman \cite{Pit(03)} and Gnedin and Pitman \cite{Gne(06)} for a detailed account on \eqref{eq:pred_gibbs} and \eqref{eq:weight}. Hereafter we briefly recall two noteworthy examples of Gibbs-type random probability measures: the two parameter Poisson--Dirichlet process and the normalized generalized Gamma process. 

\begin{exe}\label{ex_pd}
Let $(a)_{n}$ be the rising factorial of $a$ of order $n$, i.e. $(a)_{n}=\prod_{0\leq i\leq n-1}(a+i)$, for $a>0$. For any $\alpha\in(0,1)$ and $\theta>-\alpha$ the two parameter Poisson--Dirichlet process, say $P_{\alpha,\theta}$, is a Gibbs-type random probability measure with
\begin{equation}\label{h_2pd}
h(t)=\frac{\alpha\Gamma(\theta)}{\Gamma(\theta/\alpha)}t^{-\theta}
\end{equation}
such that
\begin{equation}\label{eq:pred_pd}
V_{n,k_{n}}=\frac{\prod_{i=0}^{k_{n}-1}(\theta+i\alpha)}{(\theta)_{n}}.
\end{equation}
The normalized $\alpha$-stable process is $P_{\alpha,0}$, whereas the Dirichlet process may be recovered as a limiting special case for $\alpha\rightarrow0$. See, e.g., Perman et al. \cite{Per(92)}, Pitman and Yor \cite{Pit(97)}, James \cite{Jam(02)}, Pitman \cite{Pit(03)} and James \cite{Jam(13)} for detailed accounts on $P_{\alpha,\theta}$.
\end{exe}

\begin{exe}\label{ex_gg}
Let $\Gamma(\cdot,\cdot)$ be the incomplete Gamma function, i.e., $\Gamma(a,b)=\int_{b}^{\infty}x^{a-1}\exp\{-x\}\ddr x$ for $(a,b)\in \R\times\R^+$. For any $\alpha\in(0,1)$ and $\tau\geq0$ the normalized generalized Gamma process, say $G_{\alpha,\tau}$, is a Gibbs-type random probability measure with
\begin{equation}\label{h_gg}
h(t)=\edr^{\tau^{\alpha}-\tau t}
\end{equation}
such that
\begin{equation}\label{eq:pred_gg}
V_{n,k_{n}}=\frac{\alpha^{k_{n}}\edr^{\tau}}{\Gamma(n)}\sum_{i=0}^{n-1}{n-1\choose i}(-\tau^{1/\alpha})^{i}\Gamma\left(k_{n}-\frac{i}{\alpha},\tau\right).
\end{equation}
The normalized $\alpha$-stable process coincides with $G_{\alpha,0}$, whereas $G_{1/2,\tau}$ is the normalized inverse Gaussian process. See James \cite{Jam(02)}, Pitman \cite{Pit(03)}, Lijoi et al. \cite{Lij(05)}, Lijoi \cite{Lij(07)}, Lijoi et al. \cite{Lij(08)} and James \cite{Jam(13)} for detailed accounts on $G_{\alpha,\tau}$ and applications.
\end{exe}

Within the large class of predictive probabilities of the form \eqref{eq:pred_gibbs}, those of the two parameter Poisson--Dirichlet process $P_{\alpha,\theta}$ certainly stand out for their mathematical tractability, and for having an intuitive interpretability with respect to the parameter $\alpha\in(0,1)$ and $\theta>-\alpha$. See  Zabell \cite{Zab(05)} and Bacallado et al. \cite{Bac(17)} for a description of the predictive probabilities of $P_{\alpha,\theta}$ in terms of a simple generalized P\'olya like urn scheme. These desirable features of $P_{\alpha,\theta}$ arise from the product form of the  $V_{n,k_{n}}$'s in \eqref{eq:pred_pd}, which makes the ratio $V_{n+1,k_{n}+1}/V_{n,k_{n}}$ a simple linear function of $k_{n}$, and the ratio $V_{n+1,k_{n}}/V_{n,k_{n}}$ independent of $k_{n}$. Specifically, the predictive probabilities of $P_{\alpha,\theta}$ reduce to the following
\begin{equation}\label{eq:2pd_predict2}
\text{Pr}[X_{n+1}\in\cdot\,|\,\Xn]=\frac{\theta+k_{n}\alpha}{\theta+n}\nu_{0}(\cdot)+\frac{1}{\theta+n}\sum_{i=1}^{k_{n}}(n_{i}-\alpha)\delta_{X^{\ast}_{i}}(\cdot),
\end{equation}
for $n\geq1$. The weight attached to $\nu_{0}$ in \eqref{eq:2pd_predict2} can be read as a sum of two terms with distinct asymptotic orders of magnitude: i) $\alpha\kn$, referred to as the first order term, and $\theta$, referred to as the second order term. An analogous two-term decomposition holds for the weight attached to the empirical part of \eqref{eq:2pd_predict2}. Our distinction and phrasing is formally captured by writing the weights as follows
\begin{equation}\label{eq:2ndorderdef1}
\frac{\theta+k_{n}\alpha}{\theta+n} = \frac{k_{n}\alpha}{n}+\frac{\theta}{n}+o\left(\frac{1}{n}\right)
\end{equation}
and 
\begin{equation}\label{eq:2ndorderdef2}
\frac{1}{\theta+n} = \frac{1}{n}-\frac{\theta}{n^2}+\smallo{1}{n^2},
\end{equation}
where $o$ is almost sure, recovering both contributions in a two-term asymptotic decomposition. Equations \eqref{eq:2ndorderdef1} and \eqref{eq:2ndorderdef2} lead to two large $n$ approximations of the predictive distribution displayed in \eqref{eq:2pd_predict2}: i) a first order approximation of \eqref{eq:2pd_predict2}, denoted by $\sim$, is obtained by combining \eqref{eq:2pd_predict2} with the first term on the right-hand side of \eqref{eq:2ndorderdef1} and \eqref{eq:2ndorderdef2}; ii) a second order approximation of \eqref{eq:2pd_predict2}, denoted by $\approx$, is obtained by combining \eqref{eq:2pd_predict2} with the first two terms on the right-hand side of \eqref{eq:2ndorderdef1} and \eqref{eq:2ndorderdef2}. 

Ruggiero et al. \cite{Rug(13)} and Arbel et al. \cite{Arb(15)} extended the decompositions displayed in \eqref{eq:2ndorderdef1} and \eqref{eq:2ndorderdef2} to the normalized inverse Gaussian process and the normalized generalized Gamma process, respectively, thus covering the setting described in Example \ref{ex_gg}. In the next theorem we generalize \eqref{eq:2ndorderdef1} and \eqref{eq:2ndorderdef2} to the entire class of Gibbs-type priors, that is, for any continuously differentiable function $h$ and any $\alpha\in(0,1)$ we provide a two-term asymptotic decomposition for the weights $V_{n+1,k_{n}+1}/V_{n,k_{n}}$ and $V_{n+1,k_{n}}/V_{n,k_{n}}$ of the predictive probabilities \eqref{eq:pred_gibbs}.

\begin{thm}\label{teo1}
Let $\Xn$ be a sample from $P_{\alpha,h}$ featuring  $K_{n}=k_{n}\leq n$ distinct types, labelled by $X^{\ast}_{1},\ldots,X^{\ast}_{K_{n}}$, with frequencies $(N_{1},\ldots,N_{K_{n}})=(n_{1},\ldots,n_{k_{n}})$. Assume that function $h$ is continuously differentiable and denote its derivative by $h'$. Then
\begin{equation}\label{eq:2ndorderdef1_gibbs}
\frac{V_{n+1,k_{n}+1}}{V_{n,k_{n}}}=\frac{k_{n}\alpha}{n}+\frac{\beta_{n}}{n}+o\left(\frac{1}{n}\right)
\end{equation}
and
\begin{equation}\label{eq:2ndorderdef2_gibbs}
\frac{V_{n+1,k_{n}}}{V_{n,k_{n}}}=\frac{1}{n}-\frac{\beta_{n}}{n^{2}}+o\left(\frac{1}{n^{2}}\right)
\end{equation}
 for any $n\geq1$, where $\beta_n= \varphi_h(n\kn^{-1/\alpha})$ with $\varphi_h$ being defined as  $\varphi_h(t)= -th'(t)/h(t)$. 
\end{thm}

Theorem \ref{teo1} may be applied to obtain a first and a second order approximations of the predictive probabilities of an arbitrary Gibbs-type prior $P_{\alpha,h}$. This result then contributes to a remarkable simplification in the evaluation  of \eqref{eq:pred_gibbs} for any choice of the function $h$. Besides that, Theorem \ref{teo1} highlights, for large $n$, the role of $h$ from a purely predictive perspective. In particular, according to Theorem \ref{teo1}, the function $h$ does not affect the first order term in the asymptotic decompositions \eqref{eq:2ndorderdef1_gibbs} and \eqref{eq:2ndorderdef2_gibbs}, and it is sufficient to consider a second order term in order to take into account $h$. This leads to two meaningful approximations of the predictive probabilities \eqref{eq:pred_gibbs}. In particular, by considering the sole first order term in \eqref{eq:2ndorderdef1_gibbs} and \eqref{eq:2ndorderdef2_gibbs}, one obtains the first order approximation
\begin{equation}\label{eq:2pd_predict1_approx}
\text{Pr}[X_{n+1}\in\cdot\,|\,\Xn]\sim\frac{k_{n}\alpha}{n}\nu_{0}(\cdot)+\frac{1}{n}\sum_{i=1}^{k_{n}}(n_{i}-\alpha)\delta_{X^{\ast}_{i}}(\cdot),
\end{equation}
which is the predictive of the normalized $\alpha$-stable process, i.e. $h=1$. By including the second order term in \eqref{eq:2ndorderdef1_gibbs} and \eqref{eq:2ndorderdef2_gibbs}, one obtains the second order approximation
\begin{equation}\label{eq:2pd_predict2_approx}
\text{Pr}[X_{n+1}\in\cdot\,|\,\Xn]\approx
\frac{\beta_n+k_{n}\alpha}{\beta_n+n}\nu_{0}(\cdot)+
\frac{1}{\beta_n+n}\sum_{i=1}^{k_{n}}(n_{i}-\alpha)\delta_{X^{\ast}_{i}}(\cdot),
\end{equation}
which resembles the predictive probabilities \eqref{eq:2pd_predict2} of the \twoPD $P_{\alpha,\theta}$, with the parameter $\theta$ replaced by a suitable function of $h$, $\alpha$ and the number $k_{n}$ of distinct types in the sample $\mathbf{X}_{n}$. Note that~\eqref{eq:2pd_predict2_approx} is obtained by normalizing the weights \eqref{eq:2ndorderdef1_gibbs} and \eqref{eq:2ndorderdef2_gibbs} which lead to a proper predictive distribution (the weights of~\eqref{eq:2pd_predict2_approx} sum up to one) while preserving the second order approximation since 
\begin{align*}
	\frac{\beta_n+k_{n}\alpha}{\beta_n+n} 
	=
	\frac{k_{n}\alpha}{n}+\frac{\beta_{n}}{n}+o\left(\frac{1}{n}\right)
	\quad \text{and} \quad 
	\frac{1}{\beta_n+n}
	=
	\frac{1}{n}-\frac{\beta_{n}}{n^{2}}+o\left(\frac{1}{n^{2}}\right).
\end{align*}
The predictive probabilities of any Gibbs-type prior thus admit a second order approximation, for large $n$, with an error term vanishing as $\Smallo(1/n)$. More importantly, such a second order approximation maintains the same mathematical tractability and interpretability as the predictive probability of the two parameter Poisson--Dirichlet prior.

The paper is structured as follows. In Section~\ref{sec:proof} we prove Theorem \ref{teo1} and the approximate predictive probabilities displayed in Equation \eqref{eq:2pd_predict1_approx} and Equation \eqref{eq:2pd_predict2_approx}. In Section~\ref{sec:numerical} we present a numerical illustration of our approximate predictive probabilities, thus showing their usefulness from a practical point of view; \Rcode Section~\ref{sec:posterior} describes a marginal Blackwell--MacQueen P\'olya urn posterior sampling scheme based on the proposed first order and second  approximations.
 Section~\ref{sec:disc} contains a brief discussion of our results.


\section{Proof of Theorem \ref{teo1}, Equation \eqref{eq:2pd_predict1_approx} and Equation \eqref{eq:2pd_predict2_approx}} \label{sec:proof}

Throughout this section, we will use the notation $a_n\equivalent b_n$ when $a_n/b_n\to1$ as $n\to\infty$, almost surely. The main argument of the proof consists in a Laplace approximation of the integral form for $\Vnk$ in~\eqref{eq:weight} as $n\to\infty$. This approximation basically replaces an exponentially large term in an integrand by a Gaussian kernel which matches both mean and variance of the integrand. From evaluating the Gibbs-type predictive probabilities \eqref{eq:pred_gibbs} on the whole space it is clear that we have
\begin{align}\label{eq:0discovery}
\new =  1-(n-\alpha\kn)\frac{\Vnuk}{\Vnk}.
\end{align}
Denote the integrand function of~\eqref{eq:weight} by $\fn(p,t) = t^{-\alpha \kn}p^{n-1-\kn\alpha}h(t)f_{\alpha}((1-p)t)$, and denote integration over its domain $\domainfn$ by $\iint$. Then we can write
\begin{align}\label{eq:g1}
\frac{\Vnuk}{\Vnk} = \frac{1}{n-\alpha\kn}\frac{\iint p\fn}{\iint \fn}.
\end{align}
Note that this ratio of integrals coincides with $\En(P)$, that is the expectation under the probability distribution with density proportional to $\fn$. This, combined with \eqref{eq:0discovery} provides $\newtext = \En(1-P)$. In order to apply the Laplace approximation method, write the nonnegative integrand $\fn$ in exponential form $\fn = \edr^{nl_n}$, and further define functions $g(p,t) = 1-p$ and $\tilde g(p,t) = 1$. Then
\begin{align}\label{eq:Dh_laplace}
\new = \frac{\iint g \edr^{nl_n}}{\iint \tilde g \edr^{nl_n}}.
\end{align}
The mode $\mode$ of $\fn$ (or equivalently of $l_n$) is determined by the root of the partial derivatives
\begin{align}\label{eq:derivatives_p}
& n\frac{\partial l_n(p,t)}{\partial p} = \frac{n-\alpha \kn-1}{p} - t\frac{f_\alpha^\prime(t(1-p))}{f_\alpha(t(1-p))}
\end{align}
and
\begin{align}\label{eq:derivatives_t}
& n\frac{\partial l_n(p,t)}{\partial t} =  \frac{-\alpha \kn}{t} + \frac{h^\prime(t)}{h(t)} + (1-p)\frac{f_\alpha^\prime(t(1-p))}{f_\alpha(t(1-p))},
\end{align}
where $f_\alpha^\prime$ and $h^\prime$ denote respectively the derivatives of the $\alpha$-stable density $f_\alpha$ and of the function $h$. Now consider the Laplace approximations to the numerator and the denominator of the ratio \eqref{eq:Dh_laplace} with the notations set forth in Section 6.9 of Small \cite{small2010expansions}. The exponential term is identical in both integrands of the ratio~\eqref{eq:Dh_laplace}, hence the term involving $\det \fn$, the Hessian of $\fn$, is also identical and equal to
\begin{align*}
C_n = (2\pi/n)^{2/2}(-\det \fn)^{-1/2}\edr^{nl_n\mode}.
\end{align*}
Thus it simplifies in the ratio. One needs only to consider the asymptotic series expansions, where we require a second order term $a\mode$ for the numerator, that is
\begin{align*}
\new = \frac{C_n\times\left(g\mode+\frac{1}{n}a\mode+\bigo{1}{n^2}\right)}{C_n\times\left(\tilde g\mode+\bigon\right)}.
\end{align*}
The expression of $a\mode$ is provided in Equation (6.14) of Small \cite{small2010expansions}. In our case, $a\mode = \Smallo(1/n)$, hence with $\tilde g=1$, the previous display simplifies to the following 
\begin{align}\label{eq:inter}
\new = g\mode +\smallon.
\end{align}
Let $\varphi_h(t) = -th'(t)/h(t)$. Note that, adding $(1-\pn)\times$\eqref{eq:derivatives_p} and $\tn\times$\eqref{eq:derivatives_t} we can write
\begin{align}\label{eq:1-p}
g\mode = 1-\pn = \frac{\alpha \kn +\varphi_h(\tn)}{n+\varphi_h(\tn)-1}
\end{align}
so, in view of~\eqref{eq:inter},
\begin{align}\label{eq:Dh_intermediate}
\new = \frac{\alpha \kn +\varphi_h(\tn)}{n+\varphi_h(\tn)-1} +\smallon.
\end{align}
Let $\psi(x) = (xf_\alpha^\prime(x))/(\alpha f_\alpha(x))$. By  \eqref{eq:derivatives_p}, $\psi((1-\pn)\tn) = (1-\pn)(n-\alpha\kn-1)/\alpha\pn$. By Theorem 2 in Arbel et al. \cite{Arb(15)}, $1-\pn\equivalent\alpha\kn/n$. Hence, $\psi((1-\pn)\tn)$ grows to infinity when $n\to\infty$ at the same rate  as $\kn$. But studying the variations of the $\alpha$-stable density $f_\alpha$, Nolan  \cite{nolan2003stable} shows that the only infinite limit of $\psi$ is in $0^+$ according to
\begin{align*}
\psi(x) \equivalentin{0^+} (\alpha/x)^{\frac{\alpha}{1-\alpha}} .
\end{align*}
In order that $\psi((1-\pn)\tn)$ matches with its infinite limit when $n\to\infty$, its argument $(1-\pn)\tn$ needs go to $0^+$, which yields to the following asymptotic equivalence
\begin{align*}
\kn\equivalent \psi((1-\pn)\tn) \equivalent \left(\frac{\alpha}{(1-\pn)\tn}\right)^{\frac{\alpha}{1-\alpha}},
\end{align*}
which in turn gives
\begin{align*}
\tn \equivalent \alpha\frac{\kn^{1-1/\alpha}}{1-\pn} \equivalent \alpha\frac{\kn^{1-1/\alpha}}{\alpha\kn/n} \equivalent \frac{n}{\kn^{1/\alpha}}\equivalent \Th,
\end{align*}
where the last equivalence is from \cite{Pit(03)}.
Since function $h$ is assumed to be positive and continuous differentiable, $\varphi_h(\Th)$ is \as well defined (and finite) and $\varphi_h(\tn)\equivalent\varphi_h(n \kn^{-1/\alpha})\equivalent\varphi_h(\Th)$ \as, so~\eqref{eq:Dh_intermediate} can be rewritten
\begin{align*}
\new = \frac{\alpha \kn}{n} +\frac{\beta_n}{n} +\smallon,
\end{align*}
where we set $\beta_n = \varphi_h(n \kn^{-1/\alpha})$. In other terms, to match the expression of the second order approximate predictive probability displayed in Equation \eqref{eq:2pd_predict2_approx}, we have
\begin{align*}
\new = \frac{\beta_n+k_{n}\alpha}{\beta_n+n} +\smallon.
\end{align*}
 The expression of the second weight in the predictive of the theorem  follows from \eqref{eq:0discovery}, i.e., 
\begin{multline*}
\frac{\Vnuk}{\Vnk} = \frac{1-\newtext}{n-\alpha\kn} \\
= \left(1-\frac{\alpha \kn}{n} +\frac{\beta_n}{n} +\smallon\right)\left(\frac{1}{n}+\frac{\alpha\kn}{n^2}+\smallo{\kn}{n^2}\right),\\
= \frac{1}{n}-\frac{\alpha\kn}{n^2}-\frac{\beta_n}{n^2}+\frac{\alpha\kn}{n^2}+\smallo{1}{n^2} = \frac{1}{n}-\frac{\beta_n}{n^2}+ \smallo{1}{n^2},
\end{multline*}
or, to match the expression of the second order approximate predictive of equation~\eqref{eq:2pd_predict2_approx},
\begin{align*}
\frac{\Vnuk}{\Vnk} = \frac{1}{\beta_n+n} + \smallo{1}{n^2}.
\end{align*}
%


\section{Numerical illustrations}\label{sec:numerical}

As we recalled in Example \ref{ex_pd}, the two parameter Poisson--Dirichlet process $P_{\alpha,\theta}$ is a Gibbs-type random probability measure with $\alpha\in(0,1)$ and $h(t)=t^{-\theta}\Gamma(\theta+1)/\Gamma(\theta/\alpha+1)$, for any $\theta>-\alpha$. By an application of Theorem \ref{teo1}, the predictive probabilities of $P_{\alpha,\theta}$ admit a first order approximation of the form \eqref{eq:2pd_predict1_approx} and a second order approximation of the form \eqref{eq:2pd_predict2_approx} with $\varphi_h(t) = \theta$, and such that $\beta_{n}=\theta$. Among Gibbs-type random probability measures with $\alpha\in(0,1)$, the two parameter Poisson--Dirichlet process certainly stands out for a predictive structure which admits a simple numerical evaluation. This made the two parameter Poisson--Dirichlet prior a natural candidate in several applications within the large class of Gibbs-type priors. Hereafter we present a brief numerical illustration to compare the predictive probabilities of $P_{\alpha,\theta}$ with their first and second order approximations given in terms of Equation \eqref{eq:2ndorderdef1} and Equation \eqref{eq:2ndorderdef2}. While there is no practical reason to make use our approximate predictive probabilities, because of the simple expression of \eqref{eq:2pd_predict2}, the illustration is useful to show the accuracy of our approximations. We then present the same numerical illustration for the normalized generalized Gamma process $G_{\alpha,\tau}$ of Example \ref{ex_gg}. We will see that, differently from the two parameter Poisson--Dirichlet process, the predictive probabilities of the normalized generalized Gamma process do not admits a simple numerical evaluation. This motivates the use of Theorem \ref{teo1}.

\begin{figure}
\begin{center}
\includegraphics[width=1\textwidth]{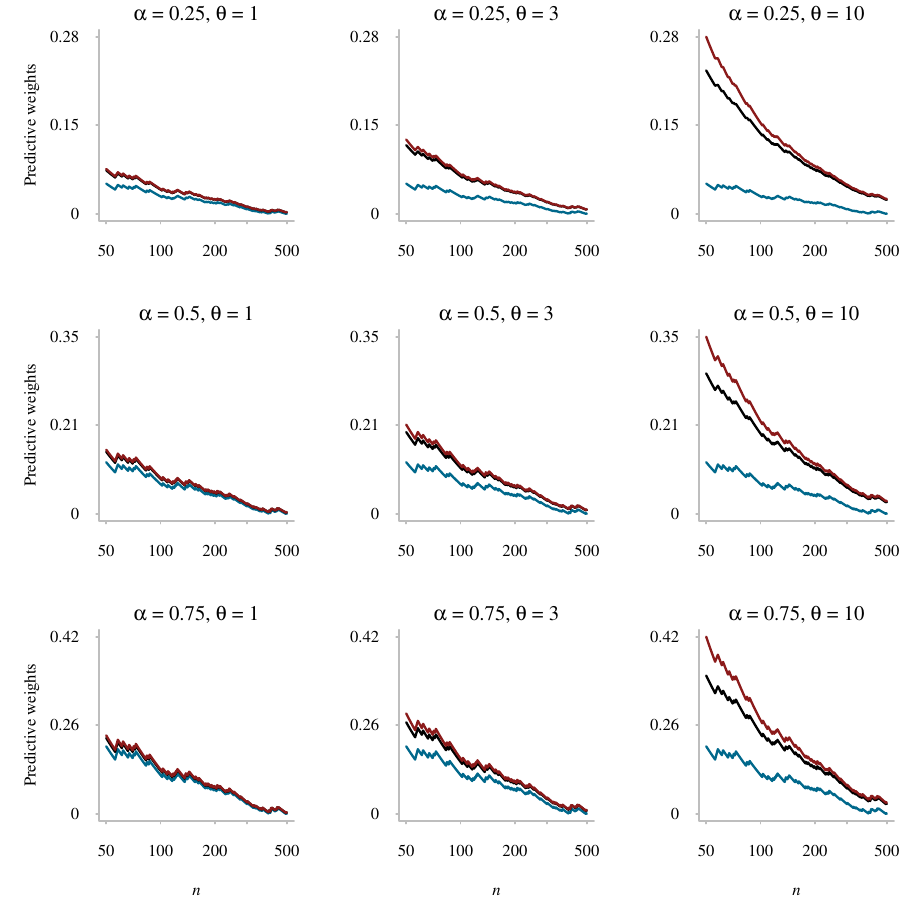}
\end{center}
\caption{Predictive weights $\newtext$  in the \twoPD. In black: the ``exact'' value~\eqref{true_pd}.  In blue: the first order approximation~\eqref{apprx1_pd}. In red: the second order approximation~\eqref{apprx2_pd}. The following values for the parameters are considered: $\alpha=0.25, 0.5$ and $0.75$ in the top, middle and bottom rows respectively; $\theta=1, 3$ and $10$ for the left, middle and right columns respectively. \samplefifty \points
\label{fig:2PD}}
\end{figure}

We consider $500$ data points sampled independently and identically distributed from the ubiquitous Zeta distribution. For any $\sigma>1$ this is a distribution with probability mass function $\text{Pr}(Z = z)\propto z^{-\sigma}$, for $z\in\mathbb{N}$. Here we choose $\sigma = 1.5$. For each $n=1,\ldots,500$ we record the number $k_{n}$ of distinct types at the $n$-th draw, and we evaluate the predictive weight $V_{n+1,k_{n}+1}/V_{n,k_{n}}$ for the two parameter Poisson--Dirichlet prior, i.e. the left-hand side of \eqref{eq:2ndorderdef1}. We consider the following pairs of parameters $(\alpha,\theta)$: $(0.25,1)$, $(0.25,3)$, $(0.25,10)$, $(0.5,1)$, $(0.5,3)$, $(0.5,10)$, $(0.75,1)$, $(0.75,3)$ and $(0.75,10)$.  For each of these pairs we compare the left-hand side of Equation \eqref{eq:2ndorderdef1} with the first term of the right-hand side of Equation \eqref{eq:2ndorderdef1} (first oder approximation) and with the first two terms of the right-hand side of Equation \eqref{eq:2ndorderdef1} (second order approximation), that are
\begin{equation}\label{true_pd}
\frac{\theta+k_{n}\alpha}{\theta+n},
\end{equation}
\begin{equation}\label{apprx1_pd}
\frac{k_{n}\alpha}{n}
\end{equation}
and
\begin{equation}\label{apprx2_pd}
\frac{k_{n}\alpha}{n}+\frac{\theta}{n},
\end{equation}
respectively.
Figure \ref{fig:2PD} shows the curve, as functions of $n$, of the ``exact" predictive weight \eqref{true_pd} and its first order approximation \eqref{apprx1_pd} and second order approximation \eqref{apprx2_pd}. The first order approximation consistently underestimates the ``exact" predictive weight, while the second order approximation consistently overestimates it. This is due to the fact that the parameter $\theta$ is positive. The discrepancy between the first order approximation and \eqref{true_pd} stays substantial even for large values of $n$, all the more for large $\theta$. On the contrary, the second order approximation consistently outperforms the first order approximation, closely following \eqref{true_pd}. For $n=500$, the ``exact" predictive weight and its second order approximation are barely distinguishable in all the considered pairs of parameters.

\subsection{The normalized generalized Gamma process}

As we recalled in Example \ref{ex_gg}, the \NGG is a Gibbs-type random probability measure with $\alpha\in(0,1)$ and $h(t)=\exp\{\tau^{\alpha}-\tau t\}$, for any $\tau\geq0$. From Theorem \ref{teo1}, the predictive probabilities of the \NGG admit a first order approximation of the form \eqref{eq:2pd_predict1_approx} and a second order approximation of the form \eqref{eq:2pd_predict2_approx} with $\varphi_h(t) = \tau t$, and
\begin{displaymath}
\beta_{n}=\frac{\tau n}{\kn^{1/\alpha}}.
\end{displaymath}
The predictive probabilities of the \NGG are of the form \eqref{eq:pred_gibbs}, with the predictive weights $V_{n+1,k_{n}+1}/V_{n,k_{n}}$ and $V_{n+1,k_{n}}/V_{n,k_{n}}$ admitting an explicit (closed-form) expression in terms of \eqref{eq:pred_gg}. However, differently from the two parameter Poisson--Dirichlet process, the evaluation of the predictive weights is cumbersome, thus preventing their practical implementation. In particular, as pointed out in Lijoi et al. \cite{Lij(07a)} in the context of mixture models with a normalized generalized Gamma prior, the evaluation of \eqref{eq:pred_gg} gives rise to severe numerical issues, even for not too large values of $n$. These  issues are mainly due to the evaluation of the incomplete gamma function, as well as with handling very small terms and very large terms within the summation \eqref{eq:pred_gg}. Because of these numerical issues in evaluating \eqref{eq:pred_gg}, we propose an alternative approach to evaluate the $V_{n,k_{n}}$'s of the \NGG. This is a Monte Carlo approach, and it relies on the fact that $V_{n,k_{n}}$ in \eqref{eq:pred_gg} can be written as the expectation of a suitable ratio of independent random variables. Recall that $f_{\alpha}$ denotes the density function of a positive $\alpha$-stable random variable. Then, using \eqref{eq:weight} with $h(t)=\exp\{\tau^{\alpha}-\tau t\}$, we can write
\begin{align}\label{expect}
\notag V_{n,\kn}&=\frac{\alpha^{\kn}}{\Gamma(n-\kn\alpha)}\int_{0}^{+\infty}\int_{0}^{1}p^{n-1-\kn\alpha}t^{-\kn\alpha}\exp\{\tau^{\alpha}-\tau t\}f_{\alpha}(t(1-p))\ddr p\ddr t\\[4pt]
&\notag=\frac{\alpha^{\kn-1}\Gamma(\kn)}{\Gamma(n)}\int_{0}^{+\infty}\exp\{\tau^{\alpha}-\tau t\}\frac{\alpha\Gamma(n)}{\Gamma(\kn)\Gamma(n-\kn\alpha)}t^{-\kn\alpha}\\
&\notag\quad\quad\quad\quad\quad\quad\quad\quad\quad\quad\quad\quad\quad\quad\quad\quad\quad\quad\int_{0}^{1}(1-p)^{n-\kn\alpha-1}f_{\alpha}(tp)\ddr p\ddr t\\[4pt]
&=\frac{\alpha^{\kn-1}\Gamma(\kn)}{\Gamma(n)}\E\left[\exp\left\{\tau^{\alpha}-\frac{\tau X}{Y}\right\}\right],
\end{align}
where $X$ and $Y$ are two independent random variables such that $Y$ is distributed according to a Beta distribution with parameter $(\kn\alpha,n-\kn\alpha)$, and $X$ is distributed according to a polynomially tilted positive $\alpha$-stable random variable, i.e., 
\begin{equation}\label{poly}
\text{Pr}[X\in \ddr x]=\frac{\Gamma(\kn\alpha+1)}{\Gamma(\kn+1)}x^{-\kn\alpha}f_{\alpha}(x)\ddr x.
\end{equation}
We refer to Pitman \cite{Pit(03)}, Pitman \cite{Pit(06)} and Devroye \cite{devroye2009random} for a detailed account on the polynomially tilted $\alpha$-stable random variable $X$. Given the representation \eqref{expect} we can perform a Monte Carlo evaluation of $V_{n,k_{n}}$ by simply sampling from the Beta random variable $Y$ and from the random variable $X$ with distribution \eqref{poly}.

Sampling form the Beta random variable $Y$ is straightforward. The random variable $X$ can be sampled by using an augmentation argument that reduces the problem of sampling $X$ to the problem of sampling a Gamma random variable and, given that, an exponentially tilted $\alpha$-stable random variable, i.e. a random variable with density function $\exp\{c^{\alpha}-cx\}f_{\alpha}(x)$, for some constant $c>0$. The problem of sampling exponentially tilted $\alpha$-stable random variables has been considered in Devroye \cite{devroye2009random} and Hofert \cite{hofert2011efficiently}. Specifically, we can write \eqref{poly} as follows
\begin{align*}
\frac{\Gamma(\kn\alpha+1)}{\Gamma(\kn+1)}x^{-\kn\alpha}f_{\alpha}(x)&=\frac{\alpha}{\Gamma(\kn)}\int_{0}^{+\infty}c^{\kn\alpha-1}\exp\{-c^{\alpha}\}\frac{\exp\{-cx\}f_{\alpha}(x)}{\exp\{-c^{\alpha}\}}\ddr c\\
&=\int_{0}^{+\infty}f_{C}(c)f_{X|C=c}(x)\ddr c,
\end{align*}
where $f_{X|C=c}$ is the density function of an exponentially tilted positive $\alpha$-stable random variable, and $f_{C}$ is the density function of the random variable $C = G^{1/\alpha}$, where $G$ being a Gamma random variable with parameter $(\kn,1)$. We apply Hofert  \cite{hofert2011efficiently} for sampling the exponentially tilted positive $\alpha$-stable random variable with density function $f_{X|C=c}$. Note that, as $\kn$ grows, the tilting parameter $C = G^{1/\alpha}$ gets larger in distribution. As a result, the acceptance probability decreases and the Monte Carlo algorithm slows down. Let Be, Ga and tSt respectively denote Beta, Gamma and exponentially tilted positive $\alpha$-stable distributions, and let $\Gamma_l$ represents the logarithm of the $\Gamma$ function. Hereafter is the step-by-step pseudocode for the Monte Carlo evaluation of the $V_{n,k_{n}}$'s: 
\begin{enumerate}
\item Set $M=10^4$, $n$, $\kn$, $\alpha$, $\tau$;
\item Sample $Y\sim\text{Be}(\alpha \kn, n-\alpha \kn)$ of size $M$;
\item Sample $G\sim \text{Ga}(\kn,1)$ of size $M$;
\item Sample $X\sim \text{tSt}(\alpha, G^{1/\alpha})$ of size $M$
\item Set $v = (\kn-1)\log\alpha +\Gamma_l(\kn) - \Gamma_l(n)+\tau^\alpha -\tau X/Y$;
\item Set $V = \exp(v)$.
\end{enumerate}

\begin{figure}
\begin{center}
	\includegraphics[width=1\textwidth]{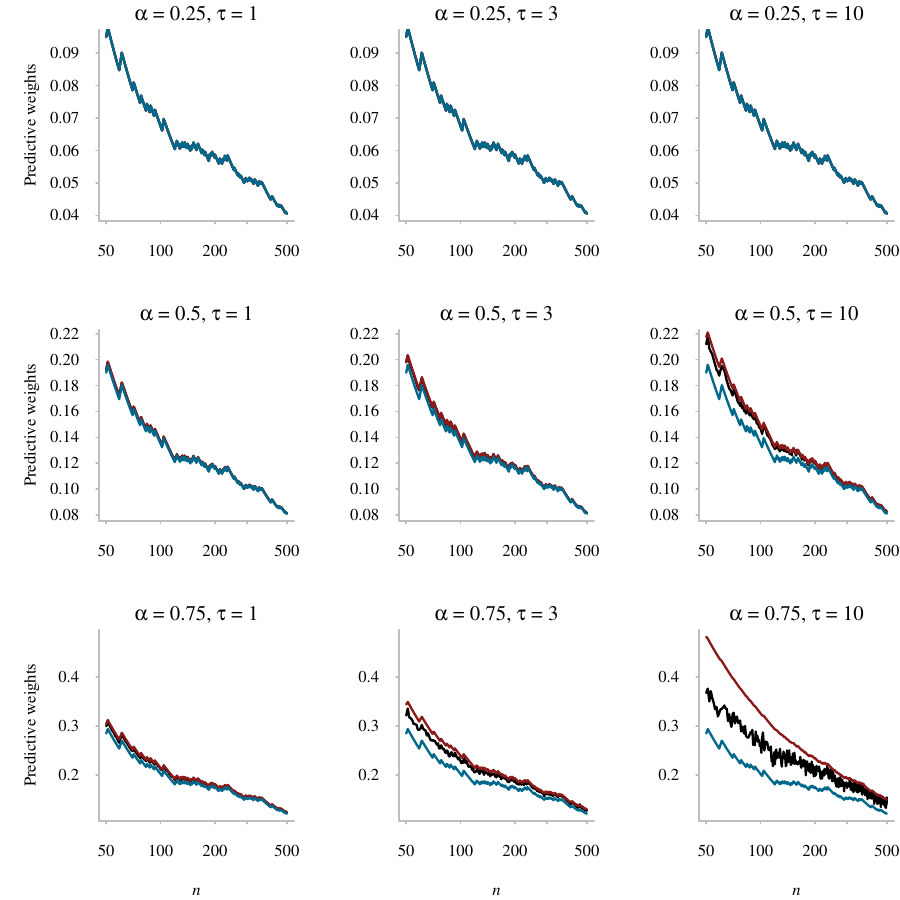}
\end{center}
\caption{
Predictive weights $\newtext$  in the \NGG. In black: the ``exact'' value evaluated by the Monte Carlo approach.  In blue: the first order approximation~\eqref{eq:approx1_NGG}. In red: the second order approximation~\eqref{eq:approx2_NGG}. The following values for the parameters are considered: $\alpha=0.25, 0.5$ and $0.75$ in the top, middle and bottom rows respectively; $\tau=1, 3$ and $10$ for the left, middle and right columns respectively. \samplefifty \points
\label{fig:NGG}
}
\end{figure}

In the same setting described for the \twoPD, we perform a numerical study for the \NGG. More specifically, $500$ data points are sampled independently and identically distributed from the Zeta distribution with parameter $\sigma=1.5$. We consider the following pairs of parameters $(\alpha,\tau)$: $(0.25,1)$, $(0.25,3)$, $(0.25,10)$, $(0.5,1)$, $(0.5,3)$, $(0.5,10)$, $(0.75,1)$, $(0.75,3)$ and $(0.75,10)$. For these pairs of parameters the predictive weight $\newtext$ is evaluated by means of the above steps 1-6, and this evaluation is compared with the first order approximation and with the second order approximation of $\newtext$ given by Theorem \ref{teo1}, i.e.
\begin{equation}\label{eq:approx1_NGG}
\frac{\kn\alpha}{n} 	
\end{equation}
and
\begin{equation}\label{eq:approx2_NGG}
\frac{\kn\alpha}{n}+\frac{\tau}{\kn^{1/\alpha}},
\end{equation}
respectively. Figure \ref{fig:NGG} shows that the Monte Carlo evaluation of $\newtext$ lays between the first order approximation and the second order approximation of $\newtext$. As $n$ moves, the difference between the resulting Monte Carlo curve and the approximate curves is imperceptible for $\alpha = 0.25$; such a difference is also very small for $\tau = 1$. Larger values of $\alpha$ and/or $\tau$ lead to larger discrepancies between the Monte Carlo curve and the approximate curves. The second order approximation is consistently closer to the Monte Carlo value than the first order approximation. In particular we observe that for $n=500$ the second order approximation and the Monte Carlo value are indistinguishable, whereas the first order approximation may still be far from the Monte Carlo value for several choices of the parameters, e.g. $(\alpha,\tau)=(0.75,3)$ and $(\alpha,\tau)=(0.75,10)$.

\begin{figure}
\begin{center}
	\includegraphics[width=.7\textwidth]{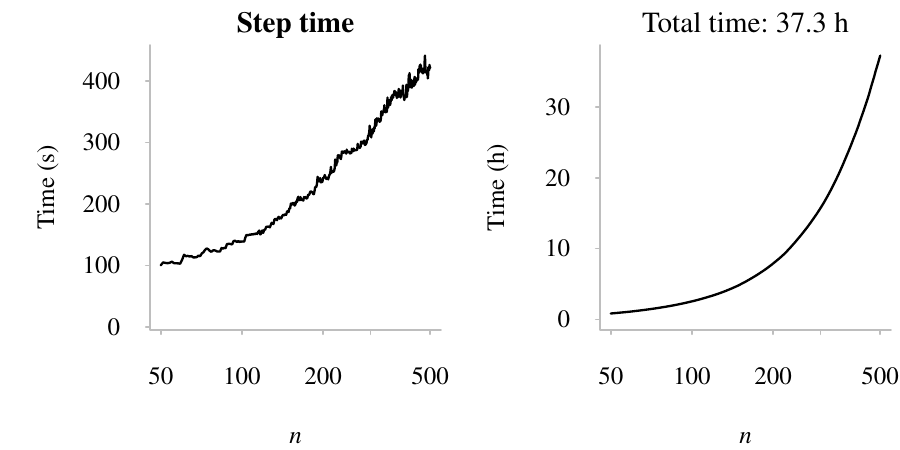}
\end{center}
\caption{
Left panel: running time (in seconds) averaged over all nine parameter configurations, and right panel: cumulated running time (in hours) averaged over all nine parameter configurations, for the Monte Carlo approach applied to the evaluation of the predictive weights $\newtext$  in the \NGG case. \samplefifty \points
\label{fig:pippo}
}
\end{figure}

\begin{figure}[h!]
\begin{center}
	\includegraphics[width=\textwidth]{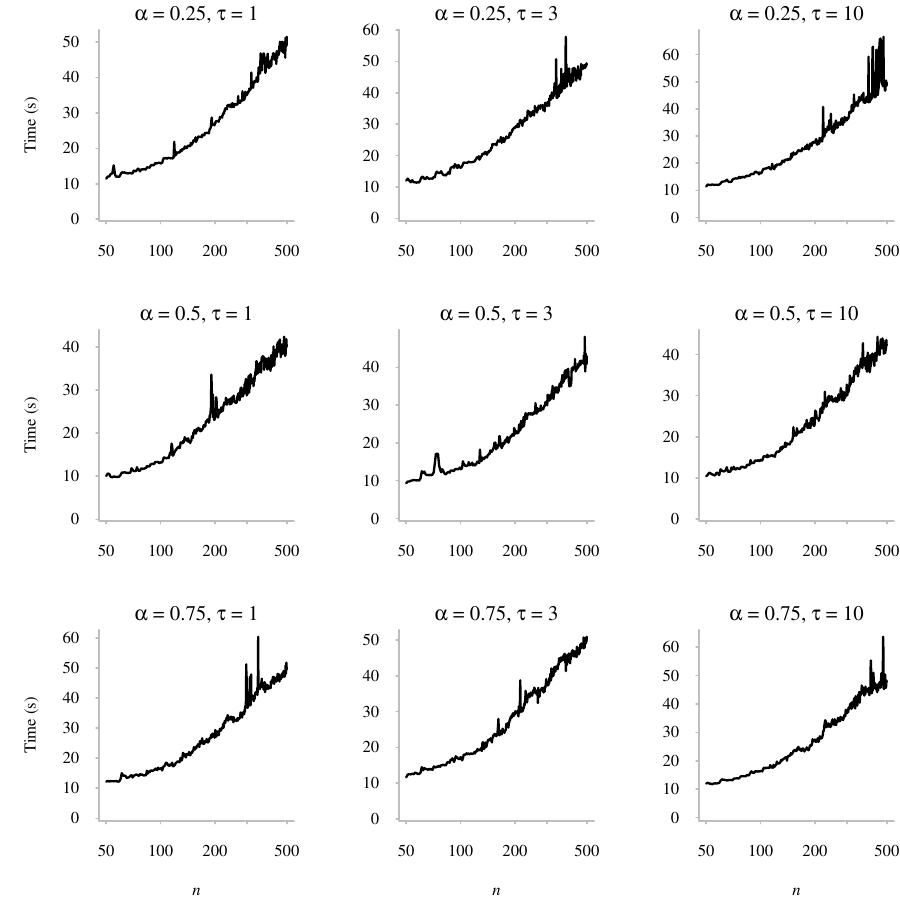}
\end{center}
\caption{
Running time (in seconds) for the Monte Carlo approach for evaluating the predictive weights $\newtext$  in the \NGG case. The following values for the parameters are considered: $\alpha=0.25, 0.5$ and $0.75$ in the top, middle and bottom rows respectively; $\tau=1, 3$ and $10$ for the left, middle and right columns respectively. \samplefifty \points
\label{fig:NGG-step-time}
}
\end{figure}

\begin{figure}
\begin{center}
	\includegraphics[width=\textwidth]{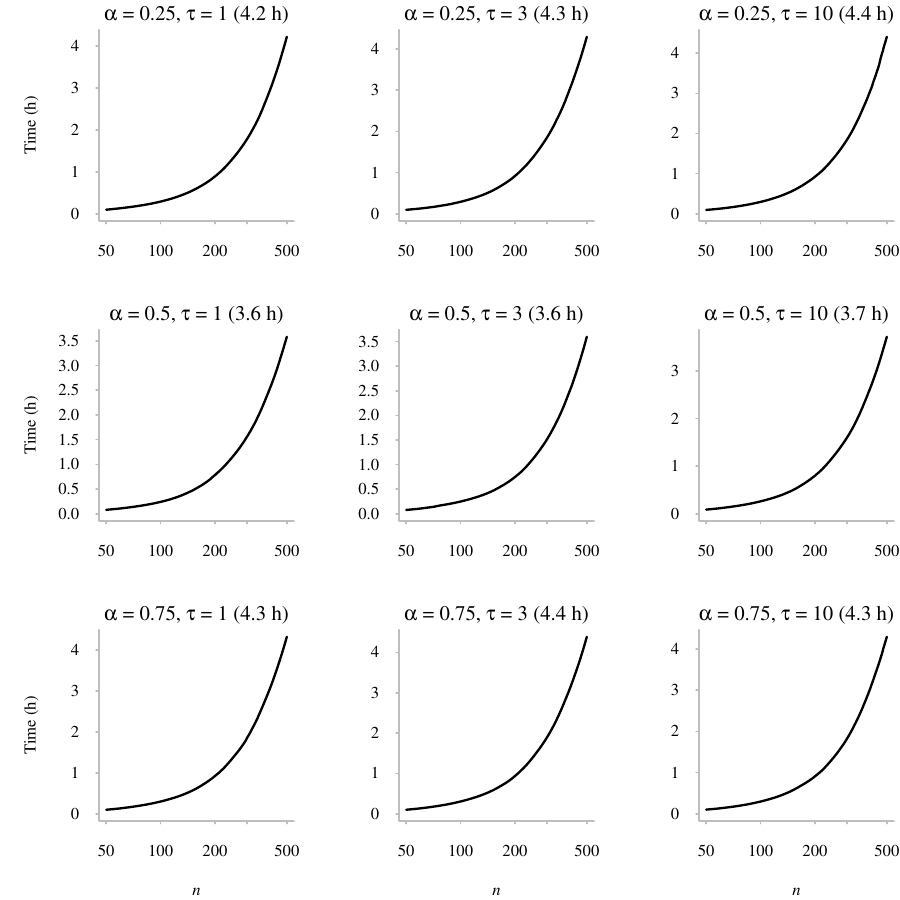}
\end{center}
\caption{
Cumulated running time (in hours) for the Monte Carlo approach for evaluating the predictive weights $\newtext$  in the \NGG case. The following values for the parameters are considered: $\alpha=0.25, 0.5$ and $0.75$ in the top, middle and bottom rows respectively; $\tau=1, 3$ and $10$ for the left, middle and right columns respectively. \samplefifty \points
\label{fig:NGG-cum-time}
}
\end{figure}

We conclude by motivating the use of the second order approximation instead of the Monte Carlo evaluation. First of all, for pairs of parameters with large $\alpha$ and large $\tau$, e.g. $(\alpha,\tau) = (0.75,10)$ in our numerical study, the Monte Carlo evaluation is extremely noisy, although we have used a large number of iterations, i.e $10^4$. In particular, as shown in Figure \ref{fig:NGG}, the noise does not vanish as $n$ grows. On the contrary, the second order approximation has a more stable behavior, and for $(\alpha,\tau) = (0.75,10)$ it converges to the bulk of the Monte Carlo curve, which makes it more reliable than the latter for large values of $n$. Furthermore, evaluating the second order approximation is fast. On the other hand, the computational burden of the Monte Carlo evaluation is very heavy, e.g. 35 hours were required for the nine configurations of Figure \ref{fig:NGG}, with $10^4$ iterations for each weight. This is because of the sampling of the exponentially tilted $\alpha$ stable random variable.  Indeed the rejection sampler originally proposed by Hofert  \cite{hofert2011efficiently} has an acceptance probability that decreases as $n$ grows, making this approach prohibitive for large sample sizes. Although our Monte Carlo code could certainly be fastened, our empirical study suggests that the computing time increases exponentially with the sample size $n$. See the average Monte Carlo running time in Figure \ref{fig:pippo}, as well as the running time and cumulated running time for each of the nine parameter configurations in Figure \ref{fig:NGG-step-time} and Figure \ref{fig:NGG-cum-time}.

\section{Posterior sampling} \label{sec:posterior}

In this section we present an application of Theorem \ref{teo1} in the context of Bayesian nonparametric mixture modeling. Among various posterior sampling schemes for Bayesian nonparametric mixture modeling, the so-called Blackwell--MacQueen P\'olya urn scheme certainly stands out. It is a Markov chain Monte Carlo sampling scheme belonging to the class of ``marginal" schemes, since it relies on the predictive distributions. See MacEachern \cite{maceachern1994estimating} and Escobar and West \cite{escobar1995bayesian} for a description of the Blackwell--MacQueen P\'olya urn scheme in the context of mixture modeling based on Dirichlet process priors, and Ishwaran and James \cite{Ish(01)} for mixture modeling based on general stick-breaking priors, e.g., the two parameter Poisson--Dirichlet process prior. We compare the Blackwell--MacQueen P\'olya urn scheme based on the exact predictive distributions with the Blackwell--MacQueen P\'olya urn scheme based on our approximated predictive distributions. The performance is evaluated by computing the Kolmogorov--Smirnov (KS) distance between the estimated distribution function and the cumulative distribution function (cdf) of the true data generating process.

As an illustrative example, we considered simulated data of varying size $n=50, 100, 200, 500$ sampled from a mixture of two Gaussian distributions, say $w_{1}\mathcal{N}(\mu_{1},\sigma_{1}^{2})+(1-w_{1})\mathcal{N}(\mu_{2},\sigma_{2}^{2})$. Precisely, we set $(\mu_{1},\sigma_{1}^{2})=(1,0.2)$, $(\mu_{2},\sigma_{2}^{2})=(10,0.2)$ and $w_{1}=0.5$. The Bayesian nonparametric mixture model can be  defined as
\begin{equation}\label{eq:mixture}
\begin{split}
&	Y_i \mid X_i \simind \mathcal{N}(Y_i\mid X_i,\sigma^2), \quad i=1,\ldots,n,\\
&	X_i \mid P_{\alpha,h} \simiid  P_{\alpha,h}, \quad i=1,\ldots,n,\\
&	 P_{\alpha,h} \sim \mathcal{P}_{\alpha,h},\\
& \sigma^2 \sim \mathcal{IG}(a,b),
\end{split}
\end{equation}
where 
$\mathcal{P}_{\alpha,h}$ denotes a Gibbs-type prior, and $\mathcal{IG}(a,b)$ stands for an inverse-gamma distribution with parameters $a$ and $b$. Following Section \ref{sec:numerical}, we focus on the two common choices for the random probability measure $P_{\alpha,h}$, namely the two parameter Poisson--Dirichlet process and the normalized generalized Gamma process. In both cases we assume that the nonatomic probability measure $\nu_{0}$ is the standard Gaussian distribution. In the model \eqref{eq:mixture} we assume that $a=b=1$.

\begin{figure}
\begin{center}
	\includegraphics[width=\textwidth]{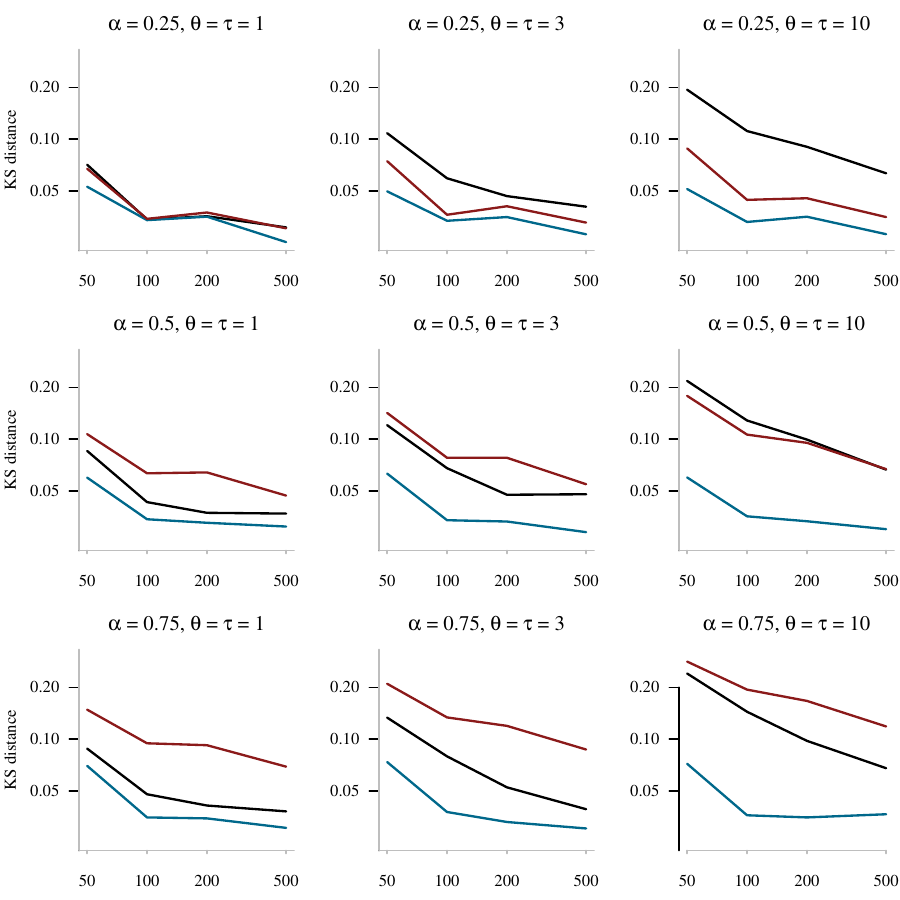}
\end{center}
\caption{
Kolmogorov--Smirnov distance between the true cdf and the cdf obtained by using the mixture model~\eqref{eq:mixture} with the following color code. 
In black: exact predictive distributions \eqref{eq:2pd_predict2} of the two parameter Poisson--Dirichlet process. In blue: first order approximation \eqref{eq:2pd_predict1_approx} of the predictive distributions of the two parameter Poisson--Dirichlet process and the normalized generalized Gamma process. In red: second order approximation \eqref{eq:2pd_predict2_approx} of the predictive distributions of the normalized generalized Gamma process. The following values for the parameters are considered: $\alpha=0.25, 0.5$ and $0.75$ in the top, middle and bottom rows respectively; $\theta = \tau=1, 3$ and $10$ for the left, middle and right columns respectively. \samplefifty \points
\label{fig:posterior}
}
\end{figure}

Under the assumption of the two parameter Poisson--Dirichlet process prior and the assumption of the normalized generalized Gamma process prior, we apply the Blackwell--MacQueen P\'olya urn scheme with the exact predictive distributions and with the corresponding approximated predictive distributions given by Theorem \ref{teo1}. We used $10^4$ iterations after a burn-in of $2\,000$. In Figure~\ref{fig:posterior}, we show the KS distance between the true distribution function and the estimated distribution function obtained by using the Blackwell--MacQueen P\'olya urn scheme with
\begin{itemize}
	\item the exact predictive distributions \eqref{eq:2pd_predict2} of the two parameter Poisson--Dirichlet process; the second order approximation \eqref{eq:2pd_predict2_approx} of the predictive distribution of the two parameter Poisson--Dirichlet process coincides with this exact predictive distribution.
	\item the first order approximation \eqref{eq:2pd_predict1_approx} of the predictive distributions of the two parameter Poisson--Dirichlet process, which coincides with the first order approximation  of the predictive distributions of the normalized generalized Gamma process.
	\item the second order approximation \eqref{eq:2pd_predict2_approx} of the predictive distributions of the normalized generalized Gamma process, which is different from the second order approximation of the predictive distribution of the two parameter Poisson--Dirichlet process.
\end{itemize}
The values of the hyperparameters $\alpha, \theta$ and $\tau$ correspond to those used in the numerical illustrations of Section~\ref{sec:numerical}. Results in Figure~\ref{fig:posterior} show  that both first and second order approximations of predictive distributions produce posterior estimates with comparable performance to that the exact predictive distribution of the two parameter Poisson--Dirichlet process.  Also, the sampling scheme based on the first order approximation outperforms the sampling scheme based on the exact predictive distributions of the two parameter Poisson--Dirichlet process, and of the second order approximation of the predictive distribution of the normalized generalized Gamma process. A reason for this superiority of the first order approximation is the following: this  first order approximation, both for the two parameter Poisson--Dirichlet process and for the normalized generalized Gamma process, boils down to the normalized $\alpha$-stable process. For a given parameter $\alpha$, such  normalized $\alpha$-stable process has a lower prior expected number of clusters than the  two parameter Poisson--Dirichlet process and the normalized generalized Gamma process counterparts. Thus the normalized $\alpha$-stable process is a better specified prior than the latter two processes for the true data generating process  which is only made of two components, leading to an overall better performance.

\section{Discussion} \label{sec:disc}

Gibbs-type priors form a flexible class of nonparametric priors, which is parameterized by an index $\alpha\in(0,1)$ and a function $h$. According to the definition of Gibbs-type random probability measures in terms of $\alpha$-stable Poisson--Kingman models, the function $h$ has the primary role of enriching the parameterization of the normalized $\alpha$-stable process by introducing additional parameters other than $\alpha$. See, e.g., Example \ref{ex_pd} and Example \ref{ex_gg}. In this paper we introduced a first order approximation \eqref{eq:2pd_predict1_approx} and a second order approximation \eqref{eq:2pd_predict2_approx} for the predictive probabilities of Gibbs-type priors, for any $\alpha\in(0,1)$ and any function $h$. In particular, we have proved that at the level of the first order approximation the function $h$ has no impact on the predictive probabilities. Indeed Equation \eqref{eq:2pd_predict1_approx} coincides with the predictive probability of the normalized $\alpha$-stable process, i.e. a Gibbs-type random probability measure with $\alpha\in(0,1)$ and $h(t)=1$. However, it is sufficient to consider a second order approximation in order to take into account the function $h$. Indeed, Equation \eqref{eq:2pd_predict2_approx} coincides with the predictive probability of the two parameter Poisson--Dirichlet process in which the parameter $\theta$ is replaced by a suitable function of $h$. The proposed approximations thus highlight the role of the function $h$ from a purely predictive perspective, and at the same time they provide practitioners with a way to easily handle the predictive probabilities of any Gibbs-type prior.


\section*{Acknowledgements} 

The authors would like to thank the Associate Editor and two anonymous Referees for their comments which helped improving substantially the paper, and Daria Bystrova for the posterior implementation of Section~\ref{sec:posterior}. This work was partly conducted during a scholar visit of Julyan Arbel at the Department of Statistics \& Data Science of the University of Texas at Austin, whose hospitality was greatly appreciated. Julyan Arbel received funding from the Grenoble Alpes Data Institute, supported by the French National Research Agency under the ``Investissements d'avenir'' program (ANR-15-IDEX-02). 
Stefano Favaro received funding from the European Research Council (ERC) under the European Union's Horizon 2020 research and innovation programme under grant agreement No 817257. Financial support from the Italian Ministry of Education, University and Research (MIUR), ``Dipartimenti di Eccellenza" grant 2018-2022, is gratefully acknowledged.



\end{document}